\documentclass[aps,amsfonts,twoside,twocolumn,amssymb,superscriptaddress]{revtex4}

\usepackage[latin1]{inputenc}
\usepackage[T1]{fontenc}
\usepackage{amsmath,amssymb,amstext,amsfonts,amsxtra}
\usepackage{amsthm}
\usepackage{graphicx}
\usepackage{mathrsfs}
\usepackage{enumerate}
\usepackage{bbm}
\usepackage{hyperref}
\usepackage{color}
\definecolor{myred}{rgb}{1,0,0}

\definecolor{myblue}{rgb}{0,0,0.8}

\definecolor{myyellow}{rgb}{0.9,0.8,0}

\definecolor{mygreen}{rgb}{0,0.6,0}

\definecolor{myorange}{rgb}{0.6,0.6,0}

\definecolor{mycerul}{rgb}{0,0.6,1}

\def\PP{{\mathbb P}}   
\def\HH{{\mathscr H}}   

\def\Qbody{{\mathcal Q}} 
\def\Cbody{{\mathcal C}} 
\def\Pbody{{\mathcal P}} 
\def\Alg{{\mathcal A}}
\def\tr{{\rm tr}}
\def\tup#1{\underline{#1}}
\def\tupxs{(\tup x|\tup s)}
\def\idty{\mathbbm{1}}
\def\ketbra#1#2{\vert #1\rangle \langle #2\vert}
\def\kettbra#1{\ketbra{#1}{#1}}              
\def\braket#1#2{\langle#1\vert#2\rangle}

\def\norm#1{||#1||}

\def\sp{{\text{span}}}

\begin{document}
\title{Extremal Quantum Correlations and Cryptographic Security}
\author{T. Franz} \email{torsten.franz@itp.uni-hannover.de} \address{Institut f\"ur Theoretische Physik, Leibniz Universit\"at Hannover\\ Appelstra\ss e 2, 30167 Hannover, Germany }
\author{F. Furrer} \address{Institut f\"ur Theoretische Physik, Leibniz Universit\"at Hannover\\ Appelstra\ss e 2, 30167 Hannover, Germany }
\author{R.~F. Werner} \address{Institut f\"ur Theoretische Physik, Leibniz Universit\"at Hannover\\ Appelstra\ss e 2, 30167 Hannover, Germany }

\begin{abstract}
We investigate a fundamental property of device independent security in quantum cryptography by characterizing probability distributions which are necessarily independent of the measurement results of any eavesdropper. We show that probability distributions that are secure in this sense are exactly the extremal quantum probability distributions. This allows us to give a characterization of security in algebraic terms. We apply the method to common examples for two-party as well as multi-party setups and present a scheme for verifying security of probability distributions with two parties, two measurement settings, and two outcomes.
\end{abstract}
\maketitle

The idea of using quantum systems for secure communication has been around for more than 25 years now. But still the boundaries of quantum cryptography have not been fully understood. Only recently a remarkable feature of quantum systems has been realized, namely that observed violations of a Bell inequality may imply  cryptographic security, even if the measurements that lead to the violation are unknown to legitimate parties. This principle goes under the name ``device independent security'' and has been proven against collective attacks \cite{Acin_Dev_ind_07}, and recently against arbitrary attacks for memoryless measurement devices \cite{Massanes_CommutinDI,Esther_2010}. But still no proof for the most general situation is known. In this paper we  focus on the question, when measurement outcomes obtained by the legitimate parties are independent of measurements performed by an eavesdropper. We give a necessary and sufficient condition for this under the assumption that the probability distributions are known without error.

We consider a quantum correlation experiment with $N$ separated parties, each performing one of $M$ different local measurements with $K$ outcomes. We denote this situation by the triple $(N,M,K)$. In a device independent scenario the parties (usually $N$=2) want to extract a secret key from the observed correlations in which the security estimation is solely based on the measured probability distributions. There are no assumption on the proper functioning of the measurement devices or the measured system, e.g., on their dimension. Probability distributions that are useful for cryptography have to feature certain properties. First, the obtained correlations should not permit a local hidden variable (LHV) model as in this case a potential adversary can have full knowledge about the correlations. Second, the correlations should only be weakly correlated to any possible measurement of an adversary. The first property is well known to be equivalent to violate a Bell inequality (see below), but the latter still lacks a concrete characterization.

In this paper we address this problem by specifying all probability distributions which do not allow a LHV model and are provably statistically independent of the knowledge of any eavesdropper. We show that these probability distributions, which we call secure, can completely be characterized in geometric terms. Indeed in the convex body $\Qbody$ of all quantum probability distributions the secure points are precisely the non-classical extremal points, i.e., those which are not deterministic and cannot be obtained as proper convex combination of other points in $\Qbody$.

The characterization of extremal points in $\Qbody$ is of general interest and numerical approaches to determine them are known \cite{Doherty_2008,BarcelonaHierarchy}. In our examples, we provide and discuss different tools to certify, respectively, find extremal probability distributions for particular $(N,M,K)$-cases. In many situations it turns out to be easier to establish a stronger property, i.e., that the algebraic structure of the measurement operators is completely determined by the probability distributions. This also leads to a stronger notion of security. The most prominent example (see example 3) are correlations which maximally violate the Clauser-Horne-Shimony-Holt (CHSH)-inequality \cite{CHSH}.

Our results have links to previous results obtained in the framework of non-signaling correlations, i.e., theories that are more general than quantum theory. One direction of our result, namely that extremality implies security was proven in \cite{Barrett_2005} for non-signalling theories in the bipartite case. In this paper, we only discuss the quantum framework, although our proofs can in principle be adapted to any non-signaling theory.

\paragraph*{Definitions.}
For simplicity, we consider the general $(N,M,K)$ case, even though the results are also valid for different numbers of measurement settings and outcomes for each party. We denote the probability for obtaining a string of outcomes $\underline x=(x_1,...,x_N)$ given a string of measurement settings $\underline s = (s_1,...,s_N)$ by $\PP (\underline x|\underline s)$. These numbers are assumed to be known exactly, i.e., we do not consider the uncertainties involved in estimating such probabilities from a finite sample.

The set of probability distributions $\PP$ conform to a LHV model, i.e., which can be realized by assuming the measurements reveal outcomes whose probabilities are predetermined, is called the set $\Cbody$ of classical correlations. It is a polytope, i.e. generated by a finite number of extremal points, which are given by the assignment of definite outcomes to each measurement. The faces (of maximal dimension) correspond to inequalities, which are linear in $\PP$, and are called (tight) Bell inequalities. In the $(2,2,2)$ case all tight Bell inequalities are equivalent to the CHSH inequality \cite{Fine_82}. A survey about Bell inequalities and further references can be found in \cite{OpenProblemBell}

We are interested in the set $\Qbody$ of quantum correlations, which is defined as the set of all probability distributions $\PP$ that can be realized by a quantum representation
\begin{equation}\label{cortabQ}
    \PP\tupxs=\tr(\rho F\tupxs),
\end{equation}
where $\rho$ is a density operator on a Hilbert space $\HH$, whose dimension is not constrained and can be infinite, and $F\tupxs=F_1(x_1|s_1)\cdots F_N(x_N|s_N)$ is a product of commuting operators on $\HH$.
{ $\{F_i(x|s)\}_x$ are the measurement operators of the observable chosen by the $i^{\rm th}$ party according to the measurement setting $s$.}
Thus the $F_i(x|s)$ are positive operators satisfying $\sum_{x=1}^KF_i(x|s)=\idty$, and have to commute for different sites, since the parties are independent. As shown in Appendix \ref{App:StandardForm}, every $\PP$ which can be realized in this way can also be realized in a simplified ``standard'' form, in which $\rho=\kettbra\Omega$ is a pure state and the operators  $F_i(x|s)$ are projections.
Moreover, in the standard form $\vert \Omega\rangle$ is cyclic for the algebra $\Alg(F)$, which is obtained from the $F_i(x|s)$ by taking products, linear combinations and limits of expectation values.
Cyclic means that the vectors $A\vert \Omega\rangle$ with $A\in\Alg(F)$ span a dense subspace in $\HH$.

The set $\Qbody$ is a closed convex set which has in contrast to $\Cbody$ a continuum of extremal points (see for instance \cite{Masanes_Qbody(222)}). Bell inequalities define the boundary between $\Cbody$ and $\Qbody$. The set $\Qbody$ can be characterized similarly by inequalities that are linear in $\PP$, satisfied by all $\PP\in\Qbody$ and tight for at least one $\PP\in\Qbody$. We call them Tsirelson inequalities. For every linear expression in $\PP$ there is a maximum on $\Cbody$, and another, usually larger one on $\Qbody$ which leads to a Tsirelson inequality. Computational methods to derive such maximal violations in $\Qbody$ are derived in \cite{BarcelonaHierarchy,Doherty_2008}. For the CHSH expression in the $(2,2,2)$-case these maxima are $2$  \cite{CHSH} and $2\sqrt2$ \cite{tsirelsson_85}, respectively. The value $4$ is achieved on the set of ``no-signalling correlations'' $\Pbody$, defined by the property that the measurement of one party does not change the probabilities observed by another. Similar to $\Cbody$, $\Pbody$ is generated by finitely many extremal points \cite{Barrett_2005}. It holds with proper inclusion $\Cbody \subset \Qbody \subset \Pbody$.

\paragraph*{Secure probability distributions.}
 We model the eavesdropper by another quantum party, whose measurements must commute with all $F\tupxs$. Accordingly, we call a probability distribution $\PP$ \emph{secure}, if $\PP$ does not factorize, i.e., $\PP\tupxs \neq \prod_{j=1}^N \PP_j(x_j|s_j)$, and for any quantum representation and any operator $E$ commuting with all $F_i(x_i|s_i)$
\begin{equation}\label{def:secure}
    \tr\Bigl(\rho E F\tupxs\Bigr)=\tr(\rho E)\,\PP\tupxs.
\end{equation}
The operator $E$ represents all possible measurements an eavesdropper could perform. The requirement that $\PP$ is not a product is necessary to exclude classical deterministic points, i.e. the extremal points of $\Cbody$, for which (\ref{def:secure}) is satisfied trivially. As we will see, this excludes all probability distributions which can be realized in LHV models.

In device independent cryptography, our definition ensures that an attack of an eavesdropper can never be better than a classical guess. The number of extractable secure bits by classical postprocessing can then be characterized by the classical smooth min-entropy \cite{classicalminentropy}.

Our first main result gives a geometric interpretation of secure probability distributions:
{\it A probability distribution $\PP$ is secure, if and only if it is extremal in $\Qbody \backslash \Cbody$}.

The argument is straightforward. Suppose, $\PP$ is secure, but not extremal. Then there exists a direct sum representation and a convex decomposition with $\PP = \lambda \PP_1 + (1-\lambda) \PP_2$, $0 \leq \lambda \leq 1$. Now use the definition (\ref{def:secure}) with $E$ being the projector onto the first/second summand to get $\PP = \PP_1$ and $\PP = \PP_2$. This shows that the convex combination is indeed trivial and $\PP$ is extremal. As all extremal correlations in $\Cbody$ are of product form, it follows that $\PP \notin \Cbody$.
Conversely, suppose $\PP$ is extremal and $\PP \notin \Cbody$. As before, we can conclude that $\PP$ cannot be of product form. Take any commuting $0<E<\idty$ and set $\lambda = \tr (\rho E)$. Define $\PP_1 = (1/\lambda) \tr (\rho E F \tupxs)$ and $\PP_2 = (1/(1-\lambda)) \tr (\rho (\idty-E) F \tupxs)$ such that $\PP = \lambda \PP_1 + (1-\lambda) \PP_2$. As $\PP$ is extremal, it holds that $\PP = \PP_1$, which is just equation (\ref{def:secure}), so $\PP$ is secure.

To decide whether a given probability distribution is secure has now been reduced to certifying extremality in $\Qbody$. This is in general a hard problem. Even in the $(2,2,2)$-case no simple algebraic constraints are known to verify extremality of a given $\PP$. In this paper, we will provide an explicit, yet limited certification scheme in example 3.

\paragraph*{Algebraically secure probability distributions.}
There is a straightforward way to strengthen the definition of secure probability distributions by extending the factorization property to a larger set of observables. The reason is that the stronger notion of security is often easier to verify.

A probability distribution $\PP$ is called \emph{algebraically secure}, if it is secure and for any quantum representation and any operator $E$ commuting with all $F_i(x_i|s_i)$
\begin{equation}\label{def:independence}
    \tr\Bigl(\rho E \tilde F \Bigr)=\tr(\rho E)\,\tr(\rho\tilde F),
\end{equation}
for all $\tilde F\in\Alg(F)$.

They are characterized as follows: {\it A probability distribution $\PP$ is algebraically secure, if and only if it is extremal in $\Qbody \backslash \Cbody$ and has a unique quantum representation, up to unitary equivalence.}

A sketch of the proof goes as follows. Assume first that $\PP$ is algebraically secure, and therefore extremal. Let $\rho=\kettbra\Omega$ together with $F_i(x_i|s_i)$, and $\rho ' = \kettbra{\Omega'}$ with $F_i'(x_i|s_i)$ be two representations of $\PP$ on suitable Hilbert spaces $\HH$, $\HH'$. Condition (\ref{def:independence}) implies that for all corresponding operators $\tilde F \in \Alg(F)$ and $\tilde F' \in \Alg(F')$, $\tr(\rho \tilde F) = \tr(\rho' \tilde F')$. Otherwise, the direct sum representation with $E$ chosen as the projector on the first or second summand contradicts (\ref{def:independence}). Define then the unitary operator $U$ via $U\tilde F \vert \Omega \rangle = \tilde F' \vert \Omega'\rangle$ which transforms one representation into the other. Because $\vert \Omega\rangle$ and $\vert\Omega'\rangle$ are cyclic $U$ can be extended to a unitary from $\HH$ to $\HH'$.
Conversely, assume that  $\PP$ is extremal and all representations are unitarily equivalent. Let $0\leq E \leq \idty$ be an operator commuting with all $F_i(x_i|s_i)$. Since $\PP$ is extremal, $\frac{1}{\tr(\rho E)} \sqrt{E}\rho \sqrt{E}$ together with the operators $F_i(x_i|s_i)$ is a valid quantum representation of $\PP$. Hence, $E=\idty$, which implies (\ref{def:independence}).

\begin{figure}[h]
 \begin{center}
    \includegraphics*[width=5cm]{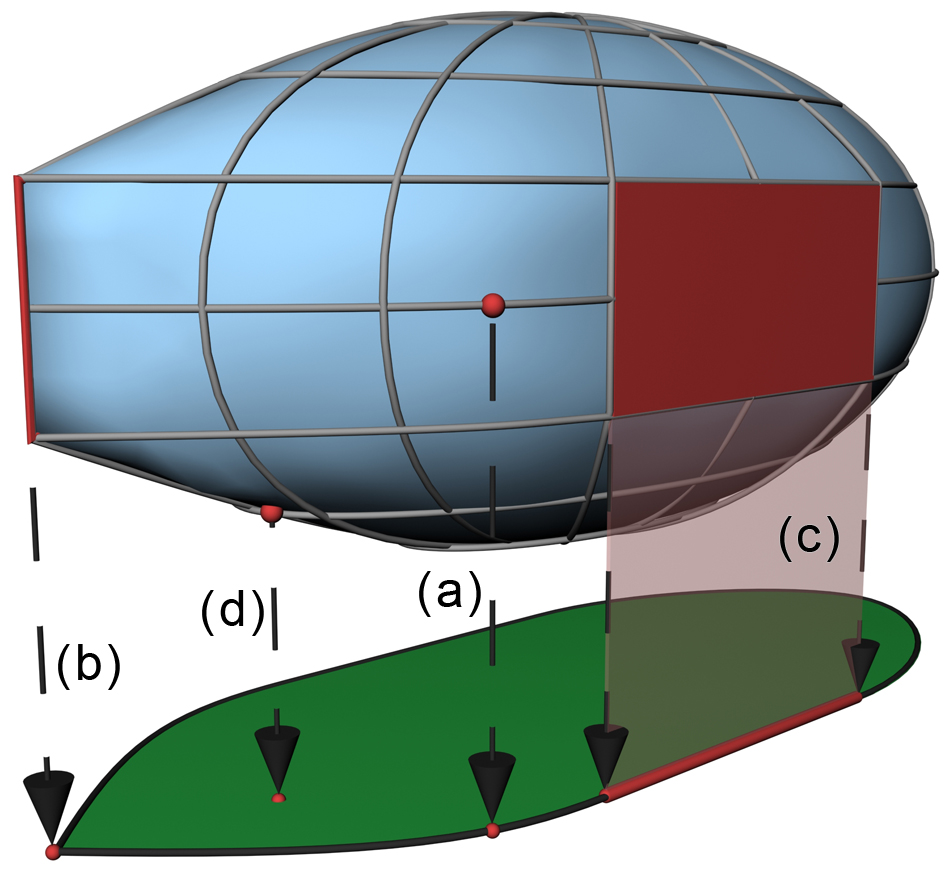}
   \caption{\label{fig:mengen} Sketch of the set of quantum representations $\mathcal S$ (above) and the set of probability distributions $\Qbody$ (below). An extremal probability distribution can either correspond to a unique point (a) or to a face of $\mathcal S$ (b).  Other faces of $\mathcal S$ can be mapped to faces of $\Qbody$ (c). Not all extremal points of $\mathcal S$ are also extremal for $\Qbody$ (d).}
\end{center}
\end{figure}

\paragraph*{Secure vs. algebraically secure.} It is now interesting to identify cases for which the notions of secure and algebraically secure coincide. To formalize the question, we can introduce a map $\Gamma$ from all possible (unitary inequivalent) quantum representations $\mathcal S$ $(=\mathcal S(N,M,K))$ to the set of probability distributions $\Qbody$. The set $\mathcal S$ can be considered as a convex set and the map $\Gamma$ is linear and surjective, but not injective. The extremal points of $\mathcal S$ are exactly the irreducible quantum representations, which are defined by the property that the only invariant subspaces of $\Alg(F)$ are $\{0\}$ and $\HH$. As shown in \cite{ChoqueBoundary}, each extremal probability distribution $\PP \in \Qbody$ admits an irreducible quantum representation. Hence, a secure probability distribution $\PP$ is algebraically secure if and only if $\Gamma^{-1}(\PP)$ is exactly one extremal point in $\mathcal S$. In FIG. \ref{fig:mengen}, the point (a) corresponds to an algebraically secure probability distribution, while the point (b) and the endpoints of the line (c) are secure, but not algebraically secure.

In the following we discuss examples, for which we provide methods to find extremal points and criterions to decide when they are also algebraically secure.

\paragraph*{Example 1: The (N,2,2)-case.}
The algebraic structure of the $(N,2,2)$-case is quite well understood (see e.g. \cite{2_Projections} and references therein). All irreducible quantum representations are in this case given on an $N$-qubit subspace $\HH= \otimes_{i=1}^{N}\mathbb C^{2}$ with an arbitrary pure state $\psi\in\HH$ and measurements, which are parameterized by $N$ angles $\theta_1,...,\theta_N$ ($\theta_i \in [0,\pi]$). The measurements are given at site $i$ as $F_i(1,1)= \frac{1}{2}(\idty + \sigma_3)$ and $F_i(1,2)=\frac{1}{2}(\idty + \sin(\theta_i)\sigma_1 +\cos(\theta_i) \sigma_3)$, together with their complements $F_i(2,s) = \idty-F_i(1,s)$. The $\sigma_i$ denote the Pauli matrices and we omitted the identities on the tensor factors for the other parties. This parametrization in $\{\theta_i\}$ and $\psi$ is sufficient to determine the whole convex body $\Qbody$. An arbitrary $\PP$ is a direct sum of at most $4^N+1$ irreducible representations. Compare \cite{Masanes_2005} for an alternative deviation of these results.

In order to find extremal points and test algebraic uniqueness we combine the above parametrization with a maximization of a Tsirelson inequality. More explicitly, for each functional given by coefficients $\{c\tupxs\}$, we can ask for the maximal quantum violation, i.e., $Q_c:= \sup_{\PP \in \Qbody} \sum_{\tup x , \tup s}c\tupxs \PP\tupxs.$
In general, $Q_c$ can be computed by a hierarchy of semi-definite programs \cite{Doherty_2008,BarcelonaHierarchy}. Here, we follow another strategy by parameterizing the corresponding operator $C = \sum c\tupxs F\tupxs=C(\theta_1,...,\theta_N)$ by means of the irreducible representations. The maximization of $\langle \psi \vert C(\theta_1,...,\theta_N)\vert \psi \rangle$ over all $\theta_i\in [0,\pi)$ and $\psi \in \mathbb{C}^{2^N}$ yields $Q_c$. Moreover, if there is exactly one set of parameters $\theta_1,...,\theta_N$ and a unique state $\psi$ for which the maximum is attained, the corresponding probability distribution $\PP$ is algebraically secure. In the case where more than one possible choice of $\theta_1,...,\theta_N,\psi$ leads to a maximal violation, we can determine the convex span of the corresponding probability distributions. This corresponds to the face given by the intersection of $\Qbody$ and the hyperplane $\{ \PP \ |\ \sum_{\tup x , \tup s}c\tupxs \PP\tupxs = Q_c\}$. Extremal points of that face are extremal points of $\Qbody$, and thus, secure probability distributions.

As a straightforward application, one can deduce that the probability distributions leading to maximal violation of Mermin's inequalities \cite{Mermin} are algebraically secure.

\paragraph*{Example 2: Certificate of extremality in the (2,2,2)-case.}
The idea of the foregoing example was to find extremal $\PP$'s by maximizing a given Tsirelson expression.
Here, we start with a particular $\PP$ and want to construct a Tsirelson inequality saturated by $\PP$. If there exists such an inequality which is not trivial, i.e. cannot be saturated by any LHV model, and no other probability distribution in $\Qbody$ saturates it (or alternatively that just one quantum representation of $\PP$ exists), extremality of $\PP$ is certified.

We focus on the (2,2,2)-case and discuss a method how to construct a maximally violated Tsirelson expressions for a given $\PP$. It comes along with a natural order of complexity for which we solve the lowest order explicitly.
The main ingredient is again the parametrization of the irreducible quantum representations by a state $\psi\in\mathbb{C}^{2} \otimes \mathbb{C}^{2}$ and two angles $\tup \theta =(\theta_A,\theta_B)$ (see previous example) for which we denote the obtained probability distribution by $\PP_{(\tup \theta,\psi)}$. Because we are only interested in extremal $\PP$s, it is sufficient to consider $\PP_{(\tup \theta,\psi)}$ with a real $\psi$ (see Appendix \ref{App:Certificate}). Since we have dichotomic measurements we can equivalently work with $\pm 1$ valued observables instead of measurement operators. We denote the observables on Alice's (Bob's) side by $A_1,A_2$ ($B_1,B_2$) and set $A_0=B_0=\idty$.

Finding a Tsirelson inequality for $\PP_{(\tup \theta,\psi)}$ is equivalent to the following task: Construct a positive operator $T=\sum_{k} P_{k}(A_i,B_j)^{\dag}P_{k}(A_i,B_j)$, with $P_{k} (A_i,B_j)$ polynomials in $A_i\otimes B_j$, $i,j=0,1,2$, such that (i) $P_{k}(A_i(\theta_A),B_j(\theta_B))\psi_0=0$ for all $k$ and (ii) $T=\sum_{i,j=0}^2 t_{ij}A_i\otimes B_j$ for all possible observables in $\HH$. Here, $A_i(\theta_A)$, $B_j(\theta_B)$ denote the observables of the representation $(\tup \theta,\psi)$.
Condition (ii) implies that $T$ can be interpreted as a linear functional of $\PP$, (i) that it is $0$ for $\PP_{(\tup \theta,\psi)}$, and the ansatz for $T$ that $T$ is a positive operator and thus its associated functional on $\PP$ is positive for each $\PP\in\Qbody$.

In order to solve the problem a constraint on the degree of the polynomials $P_k$ in the ansatz for $T$ has to be imposed. This introduces a natural hierarchy, where the order limits the possible $\PP_{(\tup \theta,\psi)}$ for which the method succeeds.
For the simplest ansatz, $P_k=\sum_{j=1}^2(\alpha_{kj}A_j\otimes\idty -\beta_{kj}\idty\otimes B_j)$ ($\alpha_{kj},\beta_{kj}\in\mathbb R$), the $\PP$ for which a Tsirelson inequality can be constructed are exactly the ones which correspond to a representation $(\phi_x^{\pm},\theta_A,\theta_B)$ with maximally entangled state $\phi_x^{\pm}= \frac{1}{\sqrt{2}}(\cos x, \mp \sin x, \sin x , \pm\cos x)$ ($x\in[0,\pi)$) for which
\begin{equation*}
\frac{\sin(2x)\sin(2x\pm \theta_B)}{\sin(2x-\theta_A)\sin(2x-\theta_A \pm \theta_B)} < 0
\end{equation*}
holds. The corresponding Tsirelson inequality and the derivation can be found in Appendix \ref{App:Certificate}.

\paragraph*{Example 3: The (2,M,2)-case for full correlations.}
The difficulty of finding extremal points in the $(2,M,2)$ scenario can be considerably reduced, as it is sufficient to consider only full correlations. This was shown by Tsirelson in \cite{tsirelsson_85} where he characterized all extremal points. In the following let $A_i$, $B_j$, $i,j \in \{1,\ldots,M\}$, denote $\pm 1$ valued observables located by Alice and Bob, and $\rho$ a density operator. The set of quantum correlations $\Qbody_{cor}$ is given by all correlation tables $c_{ij}= \tr (A_i B_j \rho)$ which can be obtained by means of a quantum representation. In \cite{tsirelsson_85} it was proven that all quantum representations of an extremal correlation table which is not deterministic have uniform marginal distributions $ \tr (A_i \rho)= \tr ( B_j \rho)=0$. Thus, non-deterministic extremal correlations in $\Qbody_{cor}$ correspond to secure probability distributions in $\Qbody$. Furthermore, an extremal correlation table which allows just one quantum representation gives rise to an algebraically secure point.

For every correlation table $c_{ij}$ exists a so-called \emph{c-system}, that is, a collection of vectors $x_i$, $y_j$ ($i,j \in \{1,\ldots,M\}$) with $\norm {x_i}\leq 1$, $ \norm {y_j}\leq 1$ in an Euclidian space with dimension $M$, such that $ c_{ij}=\langle x_i,y_j\rangle$.
If $\PP$ is extremal, the corresponding c-systems are isometric to each other, $\norm {x_i}=\norm{y_j}=1$ and the linear hull of the $\{x_i\}$ and $\{y_j\}$ coincide. Calling the dimension of the linear hull the \emph{rank} $r$ of the c-system, it further follows that $\{x_i\otimes x_i,y_j\otimes y_j\}$ span the symmetric subspace of $\mathbb R^r \otimes \mathbb R^r$. The following inequalities hold: $r \leq M$, $r \leq -1/2 + \sqrt{1/4 + 4 M)}$ and $r(r+1)/2\leq 2 M-1$. There are two cases to be distinguished. For c-systems with even rank, the representation is unique (up to unitary equivalence), while for c-systems with odd rank, there are exactly two non-equivalent representations.

With this, the question of secure versus algebraically secure is equivalent to determining the rank of the c-system which corresponds to the given correlation table. According to the inequalities above, it follows directly that all probability distributions in the $(2,2,2)$ and $(2,3,2)$-case which correspond to non-classical extremal correlations in $\Qbody_{cor}$ are algebraically secure.

\paragraph*{Acknowledgements}
T.F. acknowledges support from the DFG under grant WE-1240/12-1. F.F. acknowledges support from the Graduiertenkolleg 1463 of the Leibniz Universit\"at Hannover. R.F.W. acknowledges the support of the EU FP7 project COQUIT (contract number 233747).

\appendix

\section{Standard Form of a Quantum Representation}\label{App:StandardForm}

As described in the paper, we consider a general probability distribution obtained by $N$ parties, having $M$ measurements with $K$ outcomes each. Let $x \in \{1,..,K\}$ denote one local measurement outcome, $s \in \{1,...,M\}$ one setting and $\tup x$ ($\tup s$) denote the $N$-element strings of outcomes (measurements) for all parties. We define a probability distribution $\PP$ to be quantum and thus lying in $\Qbody$, if there exists a Hilbert space $\HH$ together with a state $\rho$ and positive operator valued measures $\{F_i(x,s)\}_{x=1}^K$, $i=1,...,N$ and $s=1,...,M$, such that $[F_i(x,s),F_j(x',s')] =0$ for every $i\neq j$, and it holds that
\begin{equation}\label{cortabQ}
    \PP\tupxs = \tr(\rho F\tupxs),
\end{equation}
where $F(\tup x, \tup s)= \prod_{i=1}^N F_i(x_i|s_i)$.

The goal is to show that for any such $\PP$ one can find a quantum representation in standard form, that is, a representation which consists of projective measurements and a pure cyclic state for $\Alg(F)$. The proof is given by an explicit construction.

First, we recapitulate the definition of $\Alg(F)$ and what it means that a vector is cyclic for $\Alg(F)$.
The algebra generated by the operators  $F_i(x,s)$, $\Alg(F)$, is defined as the closure of the set of all linear combinations of products of $F_i(x,s)$, i.e. $\sp_{\mathbb C}\{ \prod_{l=1}^m F_{i_l}(x_{i_l}|s_{i_l})| m\in \mathbb N\}\subset\mathcal{B} (\HH)$ with respect to taking expectation values (i.e., the weak* closure). This means that we add every $G\in\mathcal B(\HH)$ for which a sequence $\{G_j\}$ in $\sp_{\mathbb C}\{ \prod_{l=1}^m F_{i_l}(x_{i_l}|s_{i_l})| m\in \mathbb N\}$ exists, such that $\lim_{k\rightarrow\infty} \tr(\rho G_j)=\tr(\rho G)$ for every state $\rho$ in $\HH$. In mathematical terminology this is called a von Neumann algebra \cite{Bratteli}. We call now a pure state $\vert \Omega \rangle$ cyclic for $\Alg(F)$, if the closure of $\{G\vert\Omega\rangle | G\in \Alg(F)\}$ is the entire Hilbert space $\HH$.

We begin by turning the measurement operators $F_i(x|s)$ into projective ones, by applying a version of the Naimark dilation successively to each observable $F_i(\cdot,s)$. It suffices to do this for one of the observables, provided we verify that in this construction not only the required commutativity conditions are preserved, but also the projection valuedness of any of the other measurements. So in order to turn the observable $F_i(\cdot,s)$ in to a projective measurement, we define the Hilbert space $\widehat\HH=\bigoplus_{x=1}^K\HH_x$, where each of the $\HH_x$ is a copy of the given Hilbert space $\HH$. We denote by $P_x$ the projection onto the summand with label $x$, and introduce the isometry
$$ V:\HH\to\widehat\HH\qquad  V\phi=\bigoplus_x\sqrt{F_i(x,s)}\phi.$$
Then we will set $\widehat F_i(x,s)=P_x$, so that $V^*\widehat F_i(x,s)V=F_i(x,s)$. For other observables at the same site, e.g., $F_i(\cdot,r)$ with $r\neq s$, we set
$$ \widehat F_i(x,r)=\left\lbrace\begin{array}{cl}
                    VF_i(1,r)V^*+(\idty-VV^*) &\quad\mbox{for\ }x=1\\
                    VF_i(x,r)V^*&\quad\mbox{for\ }x>1\end{array}\right.$$
Because $V$ is an isometry, we again have $V^*\widehat F_i(x,r)V=F_i(x,r)$ for all $x$. With $V^*$ we denote the adjoint operator of $V$. Moreover, $\widehat F_i(x,r)^2=VF_i(x,r)^2V^*$ for $x>1$ and $\widehat F_i(1,r)^2=VF_i(1,r)^2V^*+(\idty-VV^*)$, so that a projective measurement remains projective. For observables at all other sites $j\neq i$ we take
$\widehat F_j(x,r)=\bigoplus_{x'} F_j(x,r)$, i.e., as the original observable acting the same on each of the summands. Once again, this preserves projective valuedness, and not only satisfies $V^*\widehat F_j(x,r)V=F_j(x,r)$, but even the stronger relation $\widehat F_j(x,r)V=VF_j(x,r)$. With this relation it is easy to see that the $\widehat F_j(x,r)$ for different $j$ (possibly $=i$) commute, so we can form the product $\widehat F\tupxs$ unambiguously, and that $V^*\widehat F\tupxs V=F\tupxs$. Hence if we define the state $\hat{\rho}=V \rho V^*$, we obtain a quantum representation of the same point $\PP\in\Qbody$, with $\widehat F_i(\cdot,s)$ projective measurements.

In order to turn $\rho$ into a pure and cyclic state we can do the Gelfand-Naimark-Segal (GNS) construction (Theorem 2.3.16 in \cite{Bratteli}) of the algbra $\Alg(F)$ with respect to the state $\rho$. We consider $\Alg(F)$ together with the positive semidefinite sesquilinear form defined by $\braket AB=\tr(\rho A^*B)$ as a pre-Hilbert space. To get a Hilbert space we first take the quotient with respect to the left ideal $I=\{A\in \Alg(F) | \tr(\rho A^*A)=0\}$ and then the completion with respect to the scalar product $\braket \cdot\cdot$. We denote the obtained Hilbert space by $\widehat\HH$ and its elements (in the densely defined subspace) are given by the equivalence classes $\psi_A=\{\tilde A | \tilde A = A + J, J\in I\}$ for $A \in \Alg(F)$. We define the representation $\pi$ of $\Alg(F)$ on $\widehat\HH$ by the equation $\pi(A)\psi_B = \psi_{AB}$ for $A,B\in\Alg(F)$. This representation is a *-homomorphism, that is, it respects products and the adjoint operation. Hence, the operators $\widehat F_i(x|s)=\pi \bigl(F_i(x|s)\bigr)$ satisfy the same commutation relation as $F_i(x|s)$ and furthermore, projections are mapped onto projections. If we set $\Omega = \psi_{\idty}$, we have that $\braket\Omega{\widehat F\tupxs\Omega}=\tr(\rho F\tupxs)=\PP\tupxs$. We therefore found a quantum representation of $\PP$ given by $\widehat F_i(x|s)$ and a pure state $\vert \Omega\rangle$ which is by definition cyclic.

\section{Lowest Order of the (2,2,2)-Certificate}\label{App:Certificate}

The goal is to check extremality for a given $\PP$ in the (2,2,2)-case. We use the same notation as introduced in the example 2 in the paper. Since we are only interested in extremal $\PP$ in $\Qbody$, we can restrict to the ones which belong to an irreducible quantum representation (see example 1 in the paper). They are described in a Hilbert space $\HH=\mathbb{C}^{2}\otimes\mathbb{C}^{2}$ and parameterized by a state $\psi\in\HH$ and angles $\theta_A,\theta_B \in [0,\pi]$, which specify the $\pm1$-valued observables $A_1,A_2$ and $B_1,B_2$ on Alice`s and Bob`s side. The concrete form of the observables are given by $A_i(\theta_A)=\sum_j t(\theta_A)_{ij} X_j$ and $B_i(\theta_B)=\sum_j t(\theta_B)_{ij} X_j$ with $X_1=\sigma_1$, $X_2=\sigma_3$ and
\begin{equation*}
t(\theta) =
\left(
  \begin{array}{cc}
    0 & 1 \\
    \sin\theta & \cos\theta \\
  \end{array}
\right).
\end{equation*}

Because the observables are all real, an extremal $\PP$ always allows a representation with a real $\psi$. To see this, we can write $\psi=\phi+i\eta$ with $\phi,\eta$ real vectors in $\HH$. Note that if $(\psi, \theta_A,\theta_B)$ generates $\PP$ so does $(\bar{\psi},\theta_A,\theta_B)$, with $\bar\psi$ the complex conjugate of $\psi$. Hence, the state $\rho=\frac{1}{2}(\vert\psi\rangle\langle\psi\vert + \vert\bar \psi\rangle\langle\bar \psi\vert)=\frac{1}{2}(\vert\phi\rangle\langle\phi\vert + \vert\eta\rangle\langle\eta\vert)$ together with $\theta_A,\theta_B$ generates the same $\PP$. But if $\PP$ is extremal then also $(\phi,\theta_A,\theta_B)$ and $(\eta,\theta_A,\theta_B)$ generates $\PP$, thus, the state can be chosen to be real.

Moreover, the case $\sin\theta=0$ corresponds to the case where the observables at Alice`s or Bob`s side commute, which corresponds to $\PP$ which can be generated by a LHV model. Hence, we restrict our attention to representations with a real $\psi$ and $\theta\neq0,\pi$.

We want to construct $T=\sum^2_{i=1} P_{i}(A_k,B_l)^{\dag}P_{i}(A_k,B_l)$ with
\begin{equation}
P_i=\sum_{j=1}^2(\alpha_{ij}A_j\otimes\idty -\beta_{ij}\idty\otimes B_j)
\end{equation}
where $\alpha$ and $\beta$ are matrices in $M_2(\mathbb{C})$, such that the conditions
\begin{description}
  \item[(i)] $P_{i}(A_l(\theta_A),B_j(\theta_B))\psi=0$ for $i=1,2$
  \item[(ii)] $T=\sum_{i,j=0}^2 c_{ij}A_i\otimes B_j$ for all possible observables $A_i,B_j$ in $\HH$
\end{description}
are satisfied. Note that a possible observable $A_i$ has to satisfy $A_i^*=A_i$ and $A_i^2=\idty$.

The restricted form of $P_i$ limits the possible irreducible representation for which the method applies, which means that it is not always possible to find coefficients $\alpha,\beta$ such that condition (i) and (ii) are satisfied. The goal is to determine for which representations this can be done and derive the corresponding Tsirelson inequality.

We start by analyzing condition (i). Using the particular form of the observables $A_i(\theta_A)$ and $B_j(\theta_B)$ expressed through $t(\theta)$, we find that $P_i\psi=0$, $i=1,2$, results in
\begin{equation}\label{cond1}
[X_i\otimes \idty] \psi = \sum_j \eta_{ij}[\idty\otimes X_j] \psi  \;\; (i=1,2)
\end{equation}
where $\eta=t(\theta_A)^{-1}\alpha^{-1}\beta t(\theta_B)$. We assumed here that $\alpha$ is invertible. However, this is not a restriction since otherwise the state $\psi$ is of product form.

In the following it is convenient to use the isomorphism between $\HH$ and the Hilbert space $M_2(\mathbb{C})$ with the Hilbert-Schmidt inner product. States $\phi=(\phi_1,\phi_2,\phi_3,\phi_4)$ in $\HH$ are identified with matrices
\begin{equation*}
\hat \phi=\left(
  \begin{array}{cc}
    \phi_1 & \phi_2 \\
    \phi_3 & \phi_4 \\
  \end{array}
\right)
\end{equation*}
and $[A\otimes\idty]\phi$ (resp. $[\idty\otimes B] \phi$) can be written as $A\hat \phi$ (resp. $\hat \phi B^T$). Moreover, we have that $\phi$ is a purification of the density matrix $\rho=(\hat\phi^* \hat\phi)^T$ on $\mathbb C^2$.
Equation (\ref{cond1}) is then equivalent to
\begin{equation}\label{cond1x}
X_i\hat \psi = \sum_{j}\eta_{ij} \hat \psi X_j.
\end{equation}

The following assertion characterizes condition (i): {\it $\psi$ admits an $\eta$ such that (\ref{cond1}) is satisfied if and only if $\hat \psi^T\psi \propto\idty$. Then, it holds that
\begin{equation}\label{eta}
\eta_{ij}=\frac{1}{2}\tr(\hat\psi^{-1} X_i \hat\psi X_j).
\end{equation}
}
The proof goes as follows. First, we note that $\hat\psi$ must be invertible. This is due to the fact that otherwise the reduced state of $\psi$ given by $(\hat\psi^* \hat\psi)^T$ has determinant $0$ and is therefore a pure state. We then multiply equation (\ref{cond1x}) with $X_k\hat\psi^{-1}$ from the left to find $\sum_{j}\eta_{ij}X_kX_j = X_k\hat\psi^{-1}X_i\hat\psi$. Recalling that $\tr(X_kX_j)=2\delta_{kj}$, we can take the trace and obtain (\ref{eta}).

We turn now to the first part of the statement. Multiplication from the right of (\ref{cond1x}) with $\hat\psi^{-1}$ shows that $X_i=\sum_{j}\eta_{ij} \hat \psi X_j\hat\psi^{-1}$. Thus, we obtain that
\begin{equation*}
\tr(X_iX_k)=\sum_{j,l}\eta_{ij}\eta_{kl}\tr(X_jX_l),
\end{equation*}
from which follows that $\eta\eta^T=\idty$. On the other hand one can check that the set $G$ of all $\hat\psi$ for which there exists a $\eta$ such that (\ref{cond1}) holds and $\det(\hat\psi)=1$, describes a group together with the usual matrix multiplication. Moreover, the map $\hat\psi \mapsto \eta(\hat\psi)$ induced by (\ref{eta}) is a group homomorphism such that $\eta(\hat\psi^T)=\eta^{-1}$. From this we can then conclude that $\hat\psi^T\hat\psi\propto \idty$ is the necessary and sufficient condition to solve (\ref{cond1}).

Because condition (i) is satisfied if and only if $\frac{1}{\det\hat\psi}\hat\psi$ is an orthogonal matrix, the possible states $\psi$ are parameterized by
\begin{equation}\label{eq,ParStates}
\phi_x^{\pm}= \frac{1}{\sqrt{2}}(\cos x, \mp \sin x, \sin x , \pm\cos x)
\end{equation}
where $x\in[0,\pi)$. The state $\psi$ determines the corresponding $\eta$ uniquely through equation (\ref{eta}).

Since the reduced state of $\psi$ is equal to $(\hat\psi^* \hat\psi)^T $, it follows directly that $\phi_x^{\pm}$ is maximally entangled. From this follows also that the expectation values of all local observables $A_l$ and $B_j$ vanish.

\begin{figure}[h]
 \begin{center}
    \includegraphics*[width=7cm]{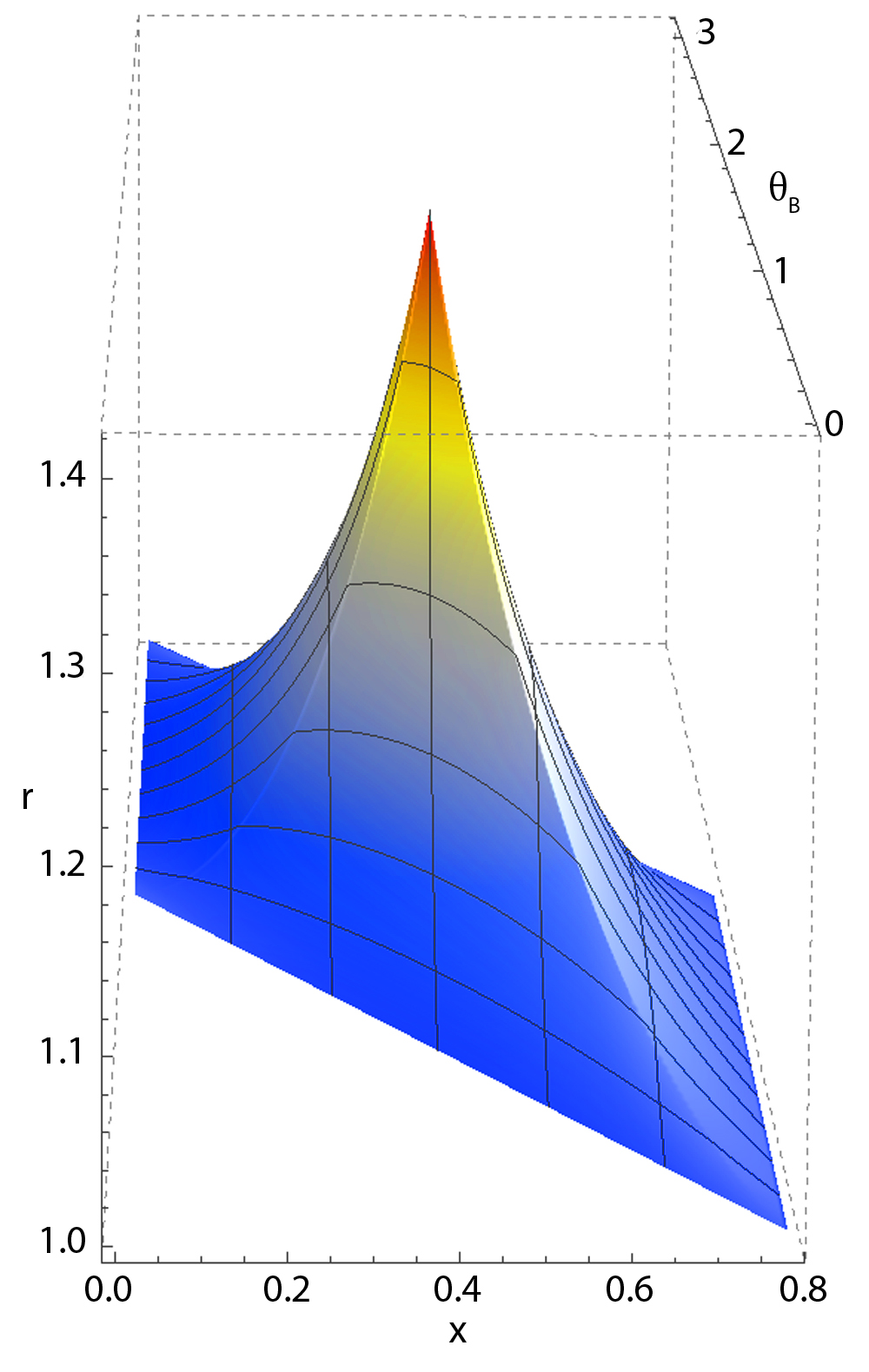}
   \caption{\label{fig:ratio} The plot shows the ratio $r$ between the maximal quantum and classical violation of the Tsirelson inequality (\ref{TsirelsonInequ}) corresponding to $(\psi=\phi_x^+, \theta_A=\frac{\pi}{2},\theta_B)$ for $\theta_B\in(0,\pi)$ and $x\in(0,\pi/4)$. The domain is restricted due to condition (\ref{cond2}). The quotient $r$ exhibits a $\pi/4$-periodic behavior in $x$ and the peak of $\sqrt2$ at $x=\pi/8$ and $\theta_B=\pi/2$ corresponds to a CHSH inequality.}
\end{center}
\end{figure}

We turn now to condition (ii) and compute the expectation value of $T$ with respect to $\phi\in\HH$,
\begin{align*}
\langle T\rangle_{\phi}
=&\tr(\alpha^*\alpha + \beta^*\beta)  \\
& - \sum_{j,k}(\alpha^*\beta +(\beta^*\alpha)^T)_{jk}\langle A_j\otimes B_k\rangle_{\phi}\\
&+\sum_{j\neq k} ( (\alpha^*\alpha)_{jk}\langle A_jA_k \rangle_{\phi} +  (\beta^*\beta)_{jk}\langle B_jB_k \rangle_{\phi}).
\end{align*}
Thus, condition (ii) requires that the matrices $\alpha^*\alpha$ and $\beta^*\beta$ are diagonal. In this case the Tsirelson inequality reads
\begin{equation}\label{TsirelsonInequ}
\sum_{j,k}(\alpha^*\beta +(\beta^*\alpha)^T)_{jk}\langle A_j\otimes B_k\rangle \leq \tr(\alpha^*\alpha + \beta^*\beta).
\end{equation}
The coefficients $c_{ij}$ in condition (ii) are therefore $c_{jk}=(\alpha^*\beta +(\beta^*\alpha)^T)_{jk}$ for $j,k=1,2$, $c_{00}=-\tr(\alpha^*\alpha + \beta^*\beta)$, and the others $0$.

Since the expectation value of $T$ is invariant under scaling and simultaneous unitary transformation of $\alpha$ and $\beta$, we can without loss of generality assume that $\alpha= \text{diag}(1,\lambda)$ with $\lambda >0$. This can always be achieved by the polar decomposition. Using now that $\eta=t(\theta_A)^{-1}\alpha^{-1}\beta t(\theta_B)$ we can write  $\beta=\alpha\gamma$ with $\gamma=t(\theta_A)\eta t(\theta_B)^{-1}$. The condition that $\beta^*\beta$ is diagonal is then equivalent to
\begin{equation}\label{cond2}
\lambda^2 = -\frac{\overline{\gamma_{11}}\gamma_{12}}{\overline{\gamma_{21}}\gamma_{22}}>0,
\end{equation}
which completely characterizes condition (ii).

Let us now summarize the results from the discussions of the two conditions. Condition (i) says that the only states $\psi$ for which the method applies are of the form (\ref{eq,ParStates}). Hence, we can constrain to representations $(\psi,\theta_A,\theta_B)$ with $\psi=\phi_x^{\pm}$. For these states we can compute $\eta$ via equation (\ref{eta}), and insert it into $\gamma=t(\theta_A)\eta t(\theta_B)^{-1}$ to find that
\begin{equation*}
\gamma_x^{\pm}=\textstyle\frac{1}{\sin\theta_B}\left(
  \begin{array}{cc}
    \textstyle\pm \sin(2x\pm\theta_B) & \textstyle\mp\sin(2x) \\
    \textstyle\pm\sin(2x-\theta_A\pm\theta_B) &\textstyle \mp\sin(2x-\theta_A) \\
  \end{array}
\right)
\end{equation*}
Condition (\ref{cond2}) can then be computed to be
\begin{equation*}
(\lambda_x^{\pm})^2 = -\frac{\sin(2x)\sin(2x\pm \theta_B)}{\sin(2x-\theta_A)\sin(2x-\theta_A \pm \theta_B)} >0.
\end{equation*}

Provided that the inequality is satisfied, the method applies and we can compute $\alpha=\text{diag}(1,\lambda)$ and $\beta=\alpha\gamma$ from which the Tsirelson inequality (\ref{TsirelsonInequ}) can be determined. Expressed in $\lambda_x^{\pm}$, one finds that
\begin{eqnarray*}
 && \textstyle\alpha^*\beta +(\beta^*\alpha)^T =  \\\\
&& \textstyle\frac{2}{\sin\theta_B}
\left(
  \begin{array}{cc}
  \scriptstyle\pm  \sin(2x\pm\theta_B) & \scriptstyle\mp\sin(2x) \\
   \scriptstyle \pm (\lambda_x^{\pm})^2\sin(2x-\theta_A\pm\theta_B) & \scriptstyle\mp (\lambda_x^{\pm})^2 \sin(2x-\theta_A) \\
  \end{array}
\right)
\end{eqnarray*}
and
\begin{equation*}
\tr(\alpha^*\alpha + \beta^*\beta)=-\frac{2\sin\theta_A\sin(4x-\theta_A\pm\theta_B)}{\sin(2x-\theta_A)\sin(2x-\theta_A\pm\theta_B)}.
\end{equation*}

Among the possible $\PP$ for which the method applies are the probability distributions which lead to maximal violation of a CHSH inequality. The corresponding representations are given by $\theta_A=\theta_B=\pi/2$ and $\psi=\phi^{\pm}_x$ with $x=\pi/8+n\pi/4$ $(n=0,1,2,3)$.


\end{document}